\documentclass[runningheads]{llncs}
\usepackage[T1]{fontenc}
\usepackage{graphicx}
\usepackage{caption}
\usepackage{subcaption}
\usepackage{blindtext}
\usepackage[colorlinks=true, urlcolor=blue, linkcolor=black, citecolor=black]{hyperref}
\usepackage{booktabs}
\usepackage{xcolor}
\usepackage{amsmath}
\usepackage{amsfonts,amssymb,exscale} 
\usepackage{amsmath}
\usepackage{mathtools}
\usepackage{bm,upgreek}
\usepackage{multirow}
\usepackage[english]{babel}
\usepackage{psfrag}
\usepackage{hyphenat}
\usepackage{mathptmx}
\newcommand{\beq}{\begin{equation}}
\newcommand{\eeq}{\end{equation}}
\newcommand{\D}  {\displaystyle}

\def\mat   #1{\mbox{\bf #1}{}}
\def\scas  #1{{\rm{#1}}{}}

\let\vec\bm
\let\ten\bm
\begin{document}
\title{Atrial constitutive neural networks}
\titlerunning{Atrial constitutive neural networks}
%
\author{Mathias Peirlinck
\inst{1},
Kevin Linka\inst{2}, \and
Ellen Kuhl\inst{3}
}
\authorrunning{M. Peirlinck et al.}
\institute{dept. BioMechanical Engineering, Delft University of Technology, Delft, The Netherlands 
\and Institute of Applied Mechanics, RWTH Aachen, Aachen, Germany \and
dept. Mechanical Engineering, Stanford University, Stanford, CA, USA
}
\maketitle   
\begin{abstract}
This work presents a novel approach for characterizing the mechanical behavior of atrial tissue using constitutive neural networks. Based on experimental biaxial tensile test data of healthy human atria, we automatically discover the most appropriate constitutive material model, thereby overcoming the limitations of traditional, pre-defined models. This approach offers a new perspective on modeling atrial mechanics and is a significant step towards improved simulation and prediction of cardiac health.

 \newline

\end{abstract}

\section{Introduction} \label{section:introduction}
The atria, the upper chambers of the heart, play an indispensable role in coordinating the cardiac cycle by efficiently receiving blood and transferring it to the ventricles. 
Although the ventricles have traditionally received more attention due to their primary blood pumping capacities 
\cite{Peirlinck2018b,Peirlinck2018a,Augustin2021,Regazzoni2022,Willems2023,Arostica2025},
there is increasing recognition that the atria significantly influence overall cardiac performance and disease progression as well \cite{Hunter2012,Moyer2015,Land2017,Augustin2019,KateBarrows2022,Casoni2023,Gerach2023,Gonzalo2024}. 
Mechanical changes in atrial tissue — due to structural remodeling, fibrosis, or age-related alterations — can profoundly affect their capacity to modulate blood flow, leading to conditions such as atrial fibrillation and chronic heart failure \cite{Peirlinck2019,Tikenogullari2023,Mase2024}. 
Consequently, accurate mechanical characterization and modeling of atrial tissue is crucial for improving our understanding of cardiac health, guiding the development of therapeutic strategies, and informing the design of medical devices aimed at treating atrial pathologies.

\smallskip \noindent
Historically, atrial tissue was long assumed to show similar mechanical behavior to myocardial tissue.
Consequently, the constitutive models used for these tissues evolved from isotropic to anisotropic, from (quasi-)linear to non-linear, and from strain-based to invariant-based models \cite{Martonova2024}. 
Over the past two decades, the evolution towards microstructurally informed constitutive models has led to increasingly more tissue-specific constitutive models for atrial tissue.
The common practice towards constitutive modeling a priori assumes a user-designed constitutive model, followed by a calibration of the associated material parameters that best fit the studied experimental tissue testing data.
Hence, atrial tissue has been modeled using transversally iostropic Fung-type strain-based constitutive models \cite{Bellini2012,Bellini2012a,KateBarrows2022}, or transversally isotropic mixed Demiray-Gasser invariant-based constitutive models \cite{Demiray1972,Gasser2006,Augustin2019}.
Originally developed for other tissue types, these a priori constitutive model choices do not necessarily reflect the intrinsic constitutive behavior of atrial tissues. 
In this work, we adopt the paradigm of constitutive neural networks \cite{Linka2023,Peirlinck2024b,Peirlinck2024a,Linka2023b,Linka2023a} to autonomously discover the best microstructure-informed constitutive model and parameters for atrial tissue from a large library of constitutive models. 
\section{Methods}
\paragraph{Biaxial tissue testing data.}
\noindent
We leverage 
biaxial extension tests performed on human atrial tissue by Bellini et al. \cite{Bellini2012}. 
In this work, thin, square-shaped atrial tissue specimens were extracted from the anterior and posterior regions of the left and right atria, and subsequently mounted in a planar biaxial tensile testing apparatus \cite{Sacks2000}. 
Based on naked eye observations of the fibers in the studied tissue specimens, 
each sample was mounted with both orthogonal fiber families \cite{Bellini2012a} parallel to either one of the sample edge pairs.
Orthogonal tensions were applied along each edge of these samples, and the protocol consisted of five distributed $t_2$:$t_1$ tension ratios which were set to 1:0.5, 1:0.75, 1:1 (equibiaxial tension), 0.75:1, and 0.5:1 respectively. 
We digitized the Green-Lagrage strain and second Piola Kirchoff stress pairs from the original work, and used standard pull-back transformations to compute the experimentally measured $\lambda_{1},\lambda_{2}$ stretch and $\hat{P}_{1},\hat{P}_{2}$ Piola stress component pairs. 
\paragraph{Kinematics.}
For the special case of homogeneous deformation during biaxial extension, 
we apply the stretches 
$\lambda_1 \ge 1$ and  $\lambda_2 \ge 1$ 
in two orthogonal directions,
and adopt an incompressibility condition,
$\mathrm{det}(\ten{F}) = \lambda_1^2 \, \lambda_2^2 \, \lambda_3^2 = 1$,
to compute the stretch in the thickness direction,
$\lambda_3= ( \lambda_1\, \lambda_2 )^{-1} \le 1$.
Based on polarized light microscopy testing of atrial tissue \cite{Bellini2012}, we impose two orthogonal fiber families in each sample.
We assume these fiber pairs remain orthogonal during tissue testing, 
such that the deformation remains homogeneous and shear free, and the deformation gradient, 
\begin{equation}
  \ten{F} 
= {\rm{diag}} \, \{ \; \lambda_1, \lambda_2, (\lambda_1 \lambda_2)^{-1} \}
\end{equation}
remains diagonal at all times. 
Based on these assumptions, the deformation of our transversally isotropic tissue samples can be completely characterized using the two isotropic invariants \cite{Peirlinck2024a}:
\begin{equation}
\begin{aligned}
  I_1 &= [\, \ten{F}^{\scas{t}} \cdot \ten{F} \,] : \ten{I} = \lambda_1^{2} + \lambda_2^{2} + (\lambda_1 \lambda_2)^{-2} \\
  I_2 &= [ I_1^2 - [\, \ten{F}^{\scas{t}} \cdot \ten{F} \,] : [\, \ten{F}^{\scas{t}} \cdot \ten{F} \,] ] = \lambda_1^{-2} + \lambda_2^{-2} + (\lambda_1 \lambda_2)^{2}
\end{aligned}
\label{iso_inv_lambda}
\end{equation}
and the anisotropic invariants for each of the orthogonal fiber families:
\begin{equation}
\begin{aligned}
I_{4,11} &= [{\ten{F}}^{\scas{t}} \cdot {\ten{F}} ] : [\vec{n}_{1}^0 \otimes \vec{n}_{1}^0] = {\lambda_1}^2  &\qquad
I_{4,22} &= [{\ten{F}}^{\scas{t}} \cdot {\ten{F}} ] : [\vec{n}_{2}^0 \otimes \vec{n}_{2}^0] = {\lambda_2}^2 \\
I_{5,11} &= [{\ten{F}}^{\scas{t}} \cdot {\ten{F}} ]^2 : [\vec{n}_{1}^0 \otimes \vec{n}_{1}^0] = {\lambda_1}^4 &\qquad
I_{5,22} &= [{\ten{F}}^{\scas{t}} \cdot {\ten{F}} ]^2 : [\vec{n}_{2}^0 \otimes \vec{n}_{2}^0] = {\lambda_2}^4 \\
\end{aligned}
\end{equation}
where $\vec{n}_{1}^0$ and $\vec{n}_{2}^0$ represent the atrial tissue's unit vector internal fiber directions in the reference configuration.
\paragraph{Constitutive equations.}
To maintain thermodynamic consistency, we define the Helmholtz free energy $\psi$ as a function of the deformation gradient, $\psi=\psi \left( \ten{F} \right)$. 
Under the assumption of no dissipative energy losses in the material, and by rewriting the Clausius–Duhem entropy inequality \cite{Planck1897} in line with the Coleman–Noll principle \cite{Coleman1959}, we obtain the Piola stress tensor $\ten{P} = \partial \Psi / \partial \ten{F}$. 
By defining the free energy function in terms of the isotropic and anisotropic invariants discussed above $\psi = \psi \left( I_1, I_2, I_{4,11},I_{4,22}, I_{5,11}, I_{5,22} \right)$, we get:
\begin{equation}
\ten{P} 
= \frac{\partial \Psi \left( I_1, I_2, I_{4,11},I_{4,22}, I_{5,11}, I_{5,22} \right)}{\partial\ten{F}}.
\label{1PKstressesorig}   
\end{equation}
To guide the selection of constitutive material models that are polyconvex, we note that mixed products of convex functions are generally not convex. Following \cite{Hartmann2003}, we therefore focus on a special subclass of free energy functions in which the free energy function is the sum of individual polynomial polyconvex subfunctions $\psi_1$, $\psi_2$, $\psi_{4,ii}$, and $\psi_{5,ii}$, such that $\psi \left( \ten{F} \right) = \psi_1 (I_1) + \psi_2 (I_2) + \sum_{i=1,2} \psi_{4,ii} \left( I_{4,ii} \right) + \sum_{i=1,2} \psi_{5,ii} \left( I_{5,ii} \right)$.
Following this design choice, we obtain the following explicit expression for the Piola stress:
\begin{equation}
\ten{P} 
= \frac{\partial \psi}{\partial I_1} \frac{\partial I_1}{\partial\ten{F}}
+ \frac{\partial \psi}{\partial I_2} \frac{\partial I_2}{\partial\ten{F}}
+ \sum_{i=1,2} \frac{\partial \psi}{\partial I_{4,ii}} \frac{\partial I_{4,ii}}{\partial\ten{F}}
+ \sum_{i=1,2} \frac{\partial \psi}{\partial I_{5,ii}} \frac{\partial I_{5,ii}}{\partial\ten{F}}.
\label{1PKstresses}   
\end{equation}
Given the homogeneous and shear free nature of the studied biaxial tensile tests and a zero stress condition in the tissue sample's normal direction, we can derive the following analytical expressions \cite{Peirlinck2024a} for the Piola stresses $P_1$ and $P_2$ in terms of the stretches $\lambda_1$ and $\lambda_2$:
\begin{equation}
\begin{aligned}
P_{1} &= 2 \D{\left[\lambda_1 - \frac{1}{\lambda_1^2 \lambda_2^{2}} \right]} \D{\frac{\partial \psi}{\partial I_1}} 
+ 2 \D{\left[\, \lambda_1 \lambda_2^2 - \frac{1}{\lambda_1^{3}} \right]} \D{\frac{\partial \psi}{\partial I_2}}
+ \D{2} \D{\lambda_1} \D{\frac{\partial \psi}{\partial I_{4,11}}} 
+ \D{4} \D{\lambda_1^3} \D{\frac{\partial \psi}{\partial I_{5,11}}} \\
P_{2} &= 2 \D{\left[ \lambda_2 - \frac{1}{\lambda_1^2 \lambda_2^{2}} \right]} \D{\frac{\partial \psi}{\partial I_1}} 
+ 2 \D{\left[\, \lambda_1^2 \lambda_2 - \frac{1}{\lambda_2^{3}} \right]} \D{\frac{\partial \psi}{\partial I_2}}
+ 2 \D{\lambda_2} \D{\frac{\partial \psi}{\partial I_{4,22}}}
+ 4 \D{\lambda_2^3} \D{\frac{\partial \psi}{\partial I_{5,22}}}
\end{aligned}
\end{equation}
%
\begin{figure}[t]
\centering
\includegraphics[width=0.75\linewidth]{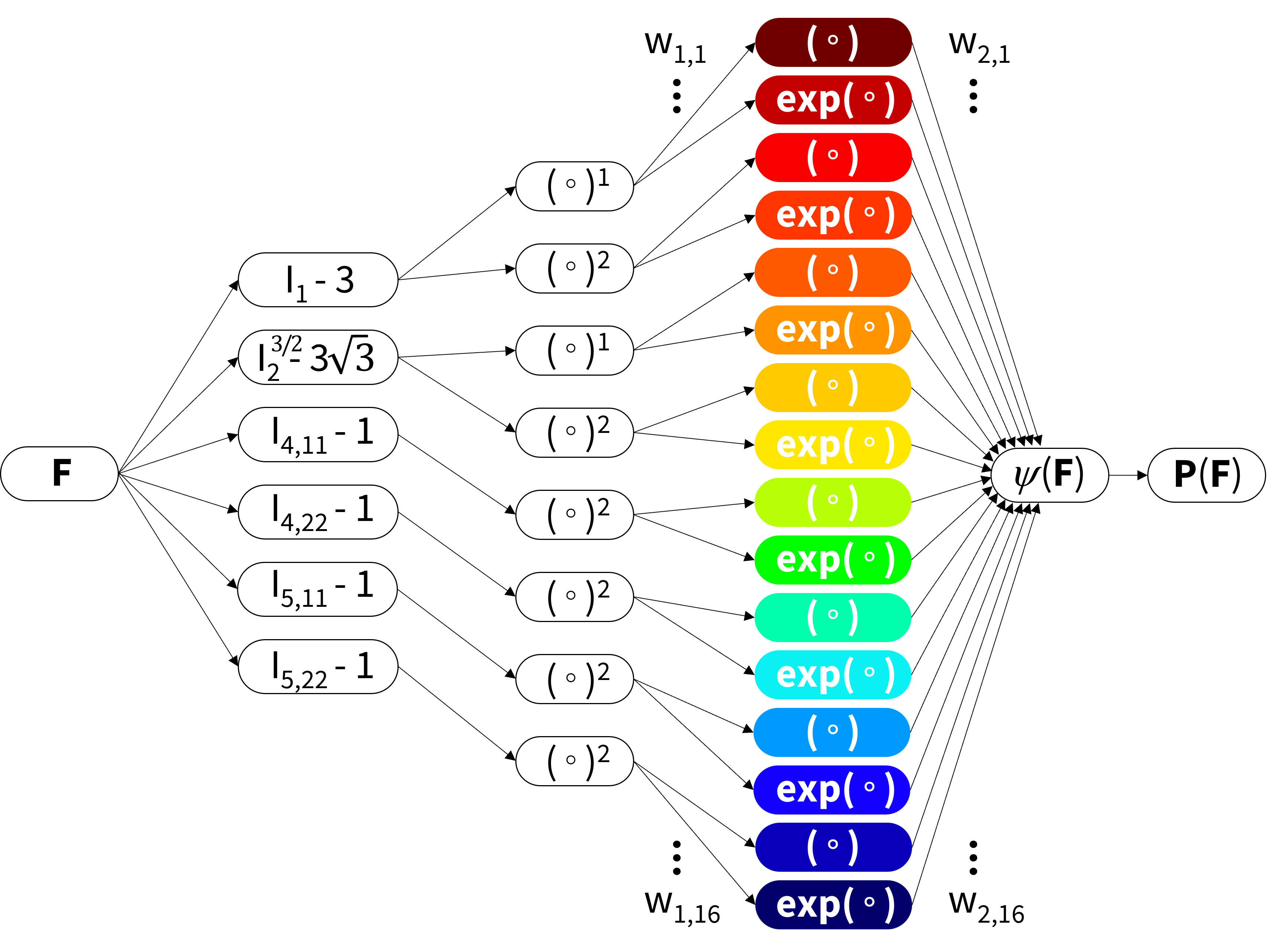}
\caption{{\bf{\sffamily{Constitutive neural network.}}} Transversely isotropic, perfectly incompressible, constitutive neural network with three hidden layers to approximate the free-energy function $\psi(I_1, I_2, I_{4,11}, I_{4,22}, I_{5,11}, I_{5,22})$ as a function of the deformation gradient $\ten{F}$ using sixteen terms. 
The zeroth layer computes the no growth corrected deformation invariants from the network input.
The first layer generates powers $(\circ)$ and $(\circ)^2$ of the zeroth and the second layer applies the identity $(\circ)$ and exponential function $(\rm{exp}(\circ))$ to these powers.}
\label{fig01}
\end{figure}%
\paragraph{Constitutive neural network.}
To discover the best material model and parameters to explain the biaxial testing data of atrial tissue, we adopt the concept of constitutive neural networks -- a special class of neural networks that satisfy the conditions of thermodynamic consistency, material objectivity, material symmetry, perfect incompressibility, polyconvexity, and physical constraints by design \cite{Linka2023}.
Figure \ref{fig01} illustrates our atrial microstructure-informed transversely isotropic, perfectly incompressible neural network with three hidden layers containing six, eight, and sixteen nodes respectively. 
The zeroth layer maps the network input, i.e. the deformation gradient, into six normalized invariant contributions: the isotropic invariants $[I_1-3]$ and $[I_2^{3/2}-3\sqrt{3}]$, and the microstructurally informed anisotropic invariants $[I_{4,11}-1]$, $[I_{4,22}-1]$, $[I_{5,11}-1]$, and $[I_{5,22}-1]$ respectively.
Here, we include normalization terms to guarantee zero free energy when there is no deformation, i.e. $\ten{F}=\ten{I}$ \cite{Linka2023,Linden2023,Kalina2023}.
The first layer generates powers $(\circ)$ and $(\circ)^2$ of these zeroth layer corrected invariants, and the second layer applies the identity $(\circ)$ and the exponential function $(\rm{exp}(\circ))$ to these powers. 
Our choice for these monotonic, continuously differentiable, smooth, and unbounded activation functions is informed by our aim to fulfill all common constitutive constrictions \cite{Linka2023,Linden2023} and polyconvexity \cite{Hartmann2003} by construction.
The total free energy function of our network takes the following explicit form, 
\beq
\begin{array}{l@{\hspace*{.10cm}}c@{\hspace*{.10cm}}l@{\hspace*{.04cm}}
              l@{\hspace*{.04cm}}l@{\hspace*{.00cm}} 
              l@{\hspace*{.04cm}}l@{\hspace*{.04cm}}l@{\hspace*{.04cm}}
              l@{\hspace*{.00cm}}l}
    \psi 
&=& w_{2,1} &w_{1,1} &[ I_1 - 3 ]
&+& w_{2,2} & [ \, \exp ( w_{1,2} & [ I_1 -3 ]&)   - 1] \\
&+& w_{2,3} &w_{1,3} &[ I_1 - 3 ]^2
&+& w_{2,4} & [ \, \exp ( w_{1,4} & [ I_1 -3 ]^2&) - 1] \\
&+& w_{2,5} &w_{1,5} &[ I_2^{3/2} -3\sqrt{3} ]
&+& w_{2,6} & [ \, \exp ( w_{1,6} & [ I_2^{3/2} -3\sqrt{3} ]&)   - 1] \\
&+& w_{2,7} &w_{1,7} &[ I_2^{3/2} -3\sqrt{3} ]^2
&+& w_{2,8} & [ \, \exp ( w_{1,8} & [ I_2^{3/2} -3\sqrt{3} ]^2&) - 1] \\
&+& w_{2,9} &w_{1,9} &[ I_{4,11} - 1 ]^2
&+& w_{2,10} & [ \, \exp ( w_{1,10} & [ I_{4,11} -1 ]^2&)   - 1] \\ 
&+& w_{2,11} &w_{1,11} &[ I_{4,22} - 1 ]^2
&+& w_{2,12} & [ \, \exp ( w_{1,12} & [ I_{4,22} -1 ]^2&) - 1] \\
&+& w_{2,13} &w_{1,13} &[ I_{5,11} - 1 ]^2
&+& w_{2,14} & [ \, \exp ( w_{1,14} & [ I_{5,11} -1 ]^2&)   - 1] \\
&+& w_{2,15} &w_{1,15} &[ I_{5,22} - 1 ]^2
&+& w_{2,16} & [ \, \exp ( w_{1,16} & [ I_{5,22} -1 ]^2&) - 1] \,,
\label{CANNenergy}
\end{array}
\eeq
Leveraging automatic differentiation in TensorFlow \cite{TensorFlow2024}, we compute the free energy function derivatives 
$\D{{\partial \psi}/{\partial I_1}}$, 
$\D{{\partial \psi}/{\partial I_2}}$,
$\D{{\partial \psi}/{\partial I_{4,11}}}$,
$\D{{\partial \psi}/{\partial I_{4,22}}}$,
$\D{{\partial \psi}/{\partial I_{5,11}}}$,
and $\D{{\partial \psi}/{\partial I_{5,22}}}$ 
which complete the definition of the computed Piola stresses in Eq. (\ref{1PKstresses}).
The network has two times sixteen weights $\mat{w}$,
which we constrain to remain non-negative, i.e. $\mat{w} \ge \mat{0}$.
We learn the network weights $\mat{w}$ by minimizing a loss function $L$ that penalizes the error between model and data. We characterize this error as the mean squared error, the $L_2$-norm of the difference between the stresses predicted by the network model,
$P_1$, $P_2$, 
and the experimentally measured stresses, 
$\hat{P}_{1}$, $\hat{P}_{2}$, 
divided by the number of training points $n_{\rm{trn}}$,
and add a penalty term, 
$\alpha \, || \mat{w} ||^p_p$, 
to allow for $L_p$ regularization,
\beq
\begin{array}{l@{\hspace*{0.1cm}}c@{\hspace*{0.1cm}}l}
L &=&  \D{\frac{1}{n_{\rm{trn}}} \sum_{j=1}^{n_{\rm{trn}}} || \, P_1 (\lambda_{1,j},\lambda_{2,j}) - \hat{P}_{1,j} \, ||^2}  \\
&+& \D{\frac{1}{n_{\rm{trn}}} \sum_{j=1}^{n_{\rm{trn}}}
|| \, P_2 (\lambda_{1,j},\lambda_{2,j}) - \hat{P}_{2,j} \, ||^2} 
 + \alpha  || \mat{w} ||^p_p
\rightarrow \mbox{min}.
\end{array}
\label{eq:loss}
\eeq
Here $\alpha \ge 0$ is a non-negative penalty parameter and 
$|| \mat{w} ||^p_p = \sum_{i=1}^{n_{\rm{par}}} |w_i|^p$ is the $L_p$ norm of the vector of the network weights $\mat{w}$. 
We train the network by minimizing the loss function (Eq. \ref{eq:loss}) using the ADAM optimizer, a robust adaptive algorithm for gradient-based first-order optimization \cite{Kingma2014}.
%
\section{Results}
\begin{figure}[h]
\centering
\includegraphics[width=1.00\linewidth]{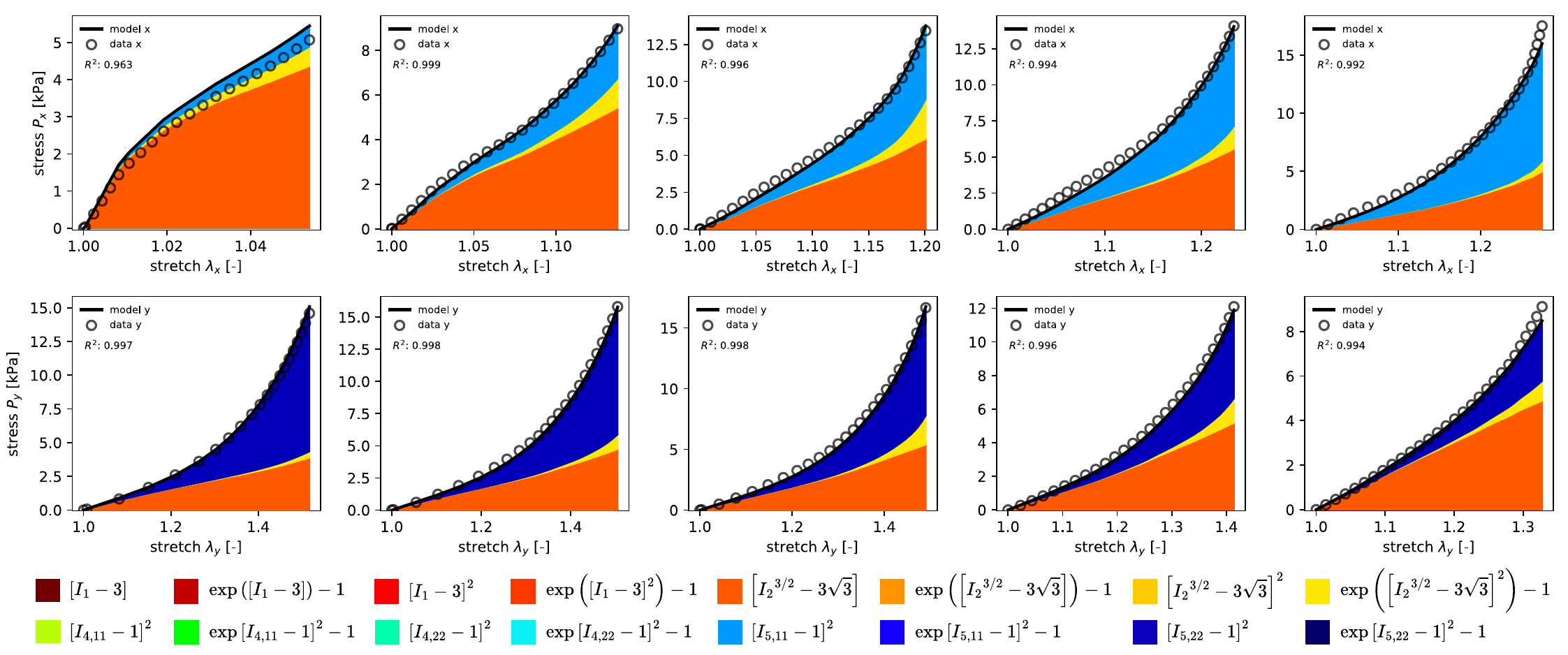}
\caption{{\bf{\sffamily{Discovered constitutive model for left atrial tissue.}}} Piola stresses $P$ as functions of stretches $\lambda$ of the constitutive neural network from Fig. \ref{fig01}, trained with all 1:0.5, 1:0.75, 1:1, 0.75:1, and 0.5:1 (left-to-right columns) $t_2$:$t_1$ tension ratios experiments on the anterior left atrial tissue sample of patient A4 simultaneously. Each individual stretch-stress curve's $R^2$ goodness of fit with respect to the original experimental data is shown in the top left corner.}
\label{fig:LAantresults}
\end{figure}

We discover the best model and parameters for both left and right atrial tissue of patient A4 \cite{Bellini2012a}. 
This task requires carefully balancing the number of terms in the discovered model with the accuracy of the fit \cite{McCulloch2024}. 
By applying $L_1$ regularization and tuning the penalty parameter over the range $\alpha = [10.0, 1.0, 0.1, 0.01]$, we learn that setting $\alpha=1.0$ strikes a desirable compromise between model sparsity and fit quality (Eq. \ref{eq:loss}).
For the anterior left atrium, we discover a four-term model, with a linear and exponential quadratic second-invariant term contribution describing the tissue's isotropic response, and two quadratic fifth-invariant terms aligning with the microstructural collagen orientations respectively:
\begin{equation}
\begin{aligned}
\psi \, &= \, \mu \left[ I_2^{3/2} - 3\sqrt{3} \right] 
+ a \left[ \exp \left( b \left[ I_2^{3/2} - 3\sqrt{3} \right]^2 \right) - 1 \right] \\
&+ \, a_1 \left[ I_{5,11} - 1 \right]^2 
+ a_2 \left[ I_{5,22} - 1 \right]^2
\end{aligned}
\end{equation}
where $\mu = 1.37$ kPa, $a = 0.0622$ kPa, $b = 0.0988$, $a_1 = 0.957$ kPa, and $a_2 = 0.394$ kPa respectively.
Figure \ref{fig:LAantresults} showcases each term's individual contribution to the overall stress response. Our discovered models showcases individual tension rate-specific $R^2$ goodness of fits ranging from 0.963 to 0.999. Our average $R^2$ goodness of fit, i.e. across all tension rations, amounts to 0.993. 
%

\begin{figure}[h]
\centering
\includegraphics[width=1.00\linewidth]{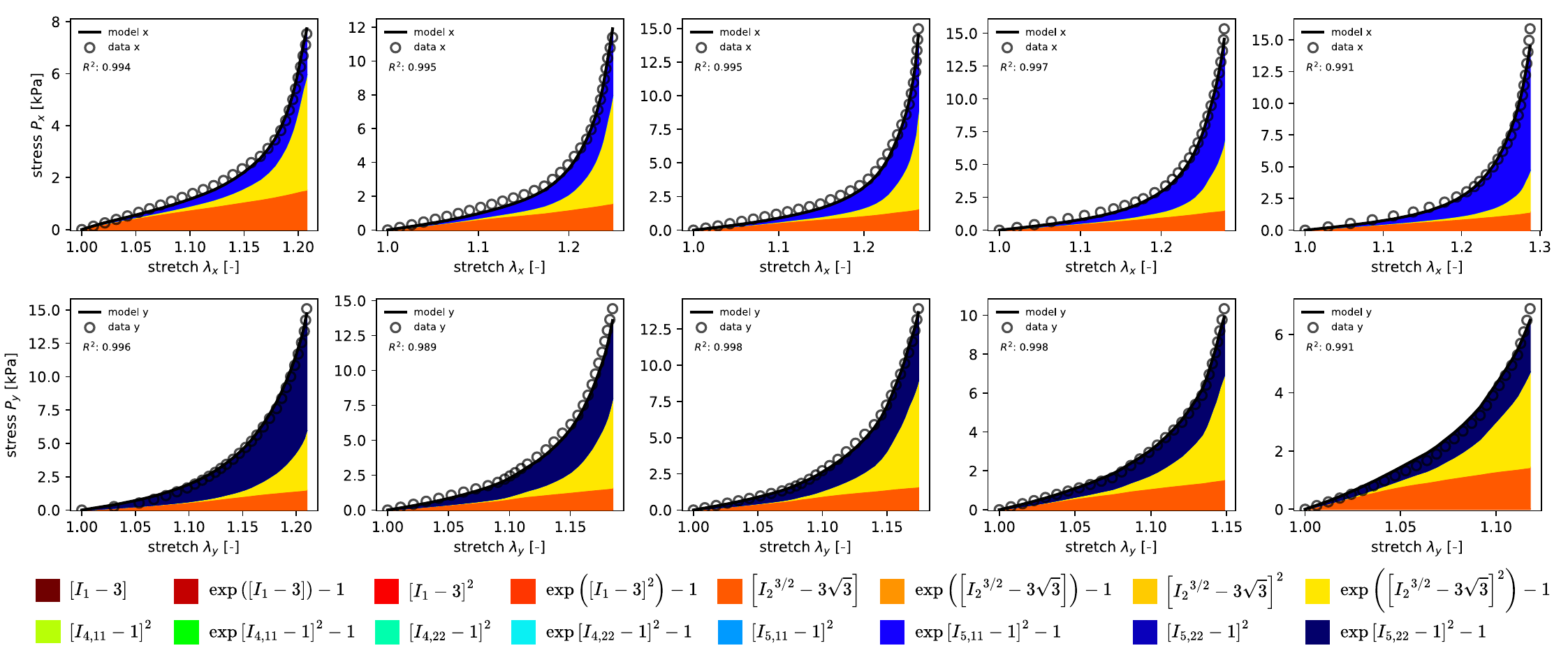}
\caption{{\bf{\sffamily{Discovered constitutive model for right atrial tissue.}}} Piola stresses $P$ as functions of stretches $\lambda$ of the constitutive neural network from Fig. \ref{fig01}, trained with all 1:0.5, 1:0.75, 1:1, 0.75:1, and 0.5:1 (left-to-right columns) $t_2$:$t_1$ tension ratios experiments on the right atrial tissue sample of patient A4 simultaneously. Each individual stretch-stress curve's $R^2$ goodness of fit with respect to the original experimental data is shown in the top left corner.}
\label{fig:RAresults}
\end{figure}

\smallskip \noindent
Figure \ref{fig:RAresults} discloses the discovered constitutive model for the right atrial tissue sample taken from patient A4 \cite{Bellini2012a}. Strikingly, out of $2^{16}-1 = 65,535$ possible models, we discover a highly similar model for both left and right atrial tissue. More specifically, our discovered four-term model features the same linear and exponential quadratic second-invariant term contribution to describe the tissue's isotropic response. Moreover, the anisotropic response of right atrial tissue is again best described using fifth-invariant terms, albeit here exponential quadratic features produce a better fit:
\begin{equation}
\begin{aligned}
\psi \, &= \, \mu \left[ I_2^{3/2} - 3\sqrt{3} \right] 
+ a \left[ \exp \left( b \left[ I_2^{3/2} - 3\sqrt{3} \right]^2 \right) - 1 \right] \\
&+ \, a_1 \left[ \exp \left( b_1 \left[ I_{5,11} - 1 \right]^2 \right) - 1 \right]
+ a_2 \left[ \exp \left( b_2 \left[ I_{5,22} - 1 \right]^2 \right) - 1 \right]
\end{aligned}
\end{equation}
Here, $\mu = 0.953$ kPa, $a = 0.0583$ kPa, $b = 0.852$, $a_1 = 0.0694$ kPa, $b_1 = 0.542$, $a_2 = 0.386$ kPa, and $b_2 = 0.498$ respectively.
Our worst and best individual $R^2$ goodness of fit amount to 0.989 and 0.998 respectively, with a total $R^2$ goodness of fit of 0.994 across all biaxial tension rations.

\section{Discussion}

This study presents a novel constitutive material model for atrial tissue that is discovered automatically using transversally isotropic microstructurally informed constitutive neural networks. 
Out of 65,535 model combinations, we consistently find the same isotropic constitutive contributions for both left and right atrial tissue featuring a linear and exponential quadratic second-invariant term \cite{Kuhl2024}. Concomitantly, both discovered models highlight the predictive value of fifth-invariant features to capture the anisotropic non-linear stiffness attributed to the two orthogonal collagen fiber families present in atrial tissue \cite{Peirlinck2024a,Vervenne2024}. 
With an average $R^2$ goodness of fit of 0.993 and 0.994 for left and right atrial tissue respectively, our work provides two highly accurate four-term constitutive models for the passive mechanical behavior of healthy human atrial tissue.

\smallskip \noindent
Given the intrinsic inter-sample variability of biological tissues, future work should focus on the validation of these discovered material models with biaxial tensile testing data taken on additional samples from the studied anatomical regions of the left atrium and the right atrium respectively.
Towards this goal, we envision a data-driven approach in which we hierarchically pool \cite{Peirlinck2019,Peirlinck2020} the biaxial tensile testing data of multiple samples and patients together towards the identification of an overarching best-fit material model, as was recently done for pulmonary arterial tissue \cite{Vervenne2024}.
In this approach, we will balance the trade-off between sample-specific $R^2$ goodness of fit metrics with an overarching $R^2$ goodness of fit metric across all samples \cite{Vervenne2024,Aggarwal2023,Linka2024,McCulloch2025}.

\smallskip \noindent
This work provides an essential reference point for understanding how atrial tissue mechanics are altered by diseases such as atrial fibrillation, and further understand how these tissue changes affect the hemodynamic loading of the ventricles, and the whole cardiovascular system throughout \cite{Aronis2019,Peirlinck2021a,Peirlinck2022,Salvador2024}.
Moreover, our discovered material model provides a important basis for the design of tissue engineered constructs that successfully take over the function of impaired regions of the atria.  
The implementation of the discovered constitutive model into finite element models \cite{Peirlinck2024c} can provide a powerful tool to assist in the planning of surgical procedures, such as the ablation of atrial tissues or the obliteration of the appendages \cite{Blackshear1996,Peirlinck2021,Banduc2025}.
%
Personalizing these atrial modeling approaches with two-stage ex vivo (biaxial tensile testing) to in vivo 
(combining non-invasive inter-atrial pressure estimation \cite{Seo2024} 
and medical imaging-based deformation analysis \cite{ArratiaLopez2023,Qayyum2024,Sillett2025,Moscoloni2025a})
mechanical stiffness calibration approaches \cite{Peirlinck2018a,Peirlinck2018b,Marx2022} and electroanatomical mapping-inferred fiber orientation maps \cite{Magana2024} will form an important stepping stone towards the modeling-supported precision medicine of the future \cite{Peirlinck2021}.

\section*{Acknowledgements}
The authors thank Matteo Salvador for fruitful discussions on atrial tissue mechanics.
M.P. acknowledges support through the NWO Veni Talent Award 20058.

%
%
%
\bibliographystyle{splncs04}
\bibliography{library}

\end{document}